\documentclass{aa}
\usepackage{epsfig}
\usepackage{amsmath}
\usepackage{txfonts}
\usepackage{natbib}
\usepackage{bm}

\begin{document}

\title{Self-consistent 3D radiative MHD simulations of coronal rain formation and evolution}
\authorrunning{Kohutova et al.}

\author{P. Kohutova\inst{1,2}, P. Antolin\inst{3}, A. Popovas\inst{1,2}, M. Szydlarski\inst{1,2} \and V. H. Hansteen\inst{1,2,4}}
\institute{Rosseland Centre for Solar Physics, University of Oslo, P.O. Box 1029, Blindern, NO-0315 Oslo, Norway\\
\email{petra.kohutova@astro.uio.no}
\and
Institute of Theoretical Astrophysics, University of Oslo, P.O. Box 1029, Blindern, NO-0315 Oslo, Norway
\and
Department of Mathematics, Physics and Electrical Engineering, Northumbria University, Newcastle Upon Tyne, NE1 8ST, UK
\and Lockheed Martin Solar and Astrophysics Laboratory, Palo Alto, CA 94304, USA\\}
\date{Received; accepted}

\abstract
{Coronal rain consists of cool and dense plasma condensations formed in coronal loops as a result of thermal instability.} 
{Previous numerical simulations of thermal instability and coronal rain formation have relied on artificially adding a coronal heating term to the energy equation. To reproduce large-scale characteristics of the corona, using more realistic coronal heating prescription is necessary.}
{We analyse coronal rain formation and evolution in a 3-dimensional radiative magnetohydrodynamic simulation spanning from convection zone to corona which is self-consistently heated by magnetic field braiding as a result of convective motions.}
{We investigate the spatial and temporal evolution of energy dissipation along coronal loops which become thermally unstable. Ohmic dissipation in the model leads to the heating events capable of inducing sufficient chromospheric evaporation into the loop to trigger thermal instability and condensation formation. The cooling of the thermally unstable plasma occurs on timescales comparable to the duration of the individual impulsive heating events. The impulsive heating has sufficient duration to trigger thermal instability in the loop but does not last long enough to lead to coronal rain limit cycles. We show that condensations can either survive and fall into the chromosphere or be destroyed by strong bursts of Joule heating associated with a magnetic reconnection events. In addition, we find that condensations can also form along open magnetic field lines.}
{We have modelled for the first time coronal rain formation in a self-consistent 3D radiative MHD simulation, in which the heating occurs mainly through the braiding and subsequent Ohmic dissipation of the magnetic field. The heating is stratified enough and lasts for long enough along specific field lines to produce the necessary chromospheric evaporation that triggers thermal instability in the corona.}


\maketitle

\section{Introduction}

Coronal rain is a common phenomenon occurring in active region coronal loops \citep{antolin_2012}. It consists of cool plasma condensations formed at coronal heights falling towards the solar surface guided by the coronal magnetic field. Coronal rain is a consequence of radiative thermal instability, which occurs when temperature gains of the plasma cannot compensate for the temperature losses \citep{parker_1953}. The resulting cooling of the plasma further increases the radiative losses, triggering runaway cooling and formation of cool and dense condensations \citep{field_1965}. In practice, this is likely to occur in a coronal loop with strong footpoint heating lasting over a time comparative to the radiative timescale of the loop \citep{johnston_2019}. The localised heating causes evaporation of chromospheric plasma into the loop, filling the upper parts of the loop with hot and dense plasma. This increase in density leads to increase in the radiative cooling rate.  As a result the overdense plasma at the top of the loop enters thermally unstable regime and local condensation occurs \citep{moschou_2015, claes_2019}. In a system that is rapidly evolving such as a coronal loop (a timescale of hours) the occurrence of thermal instability locally depends on how far away the system is from thermal equilibrium \citep{Klimchuk_2019SoPh..294..173K}. If the system is globally in a critical state of thermal equilibrium then thermal instability can occur locally due its very fast growth timescale (up to a few minutes or less) and short length scales (up to a few Mm or less) \citep[see][for a more detailed review of the process]{antolin_2020}.

The formation of coronal rain has been studied by numerical simulations in various setups focusing on different aspects of thermal instability, condensation formation and details of mass and energy transfer between the chromosphere and the corona. These include 1D hydrodynamic simulations investigating thermal stability of footpoint-heated field lines \citep{muller_2003,muller_2004, muller_2005, froment_2018}, 2.5D MHD simulations investigating coronal rain formation in an arcade magnetic field configuration highlighting the morphology of the condensations and occurrence of coronal rain limit cycles \citep{fang_2013, fang_2015}, and finally 3D MHD simulations of coronal rain formation in weak dipolar coronal magnetic fields focusing on the resulting mass drainage of the unstable coronal loop \citep{moschou_2015, xia_2017}. All numerical simulations of this phenomenon have implemented heating functions that were either constant over time \citep[e.g.][]{susino_2010}, or stochastic \citep[e.g.][]{antolin_2010}.

Thermal stability (or lack of thereof) of coronal loop is determined by the spatio-temporal characteristics of the loop heating \citep{froment_2018, johnston_2019, klimchuk_2019}. When the heating is sufficiently stratified and of high enough frequency (compared to the radiative cooling time) the loop enters a global state of thermal non-equilibrium (TNE). Such loop is unable to reach a thermal equilibrium and undergoes limit cycles of heating, in which chromospheric evaporation occurs and the loop becomes dense, and cooling, in which a runaway radiative cooling occurs and the loop depletes \citep{kuin_1982, klimchuk_2019}. During the cooling stage thermal instability can be triggered, leading to the formation of the cool condensations that appear as coronal rain \citep{antolin_2020}. Occurrence of coronal rain can therefore be used as a proxy for coronal heating \citep{antolin_2010}.

The observational evidence suggests that a significant fraction of coronal loops are in fact in a global state of thermal non-equilibrium, undergoing heating and cooling phases. This behaviour manifests as quasi-periodic intensity pulsations in EUV wavelengths that can last several days \citep{auchere_2014, froment_2015}. Such thermal non-equilibrium cycles are often accompanied by coronal rain formation \citep{auchere_2018, froment_2020}. Multiple occurrences of coronal rain in the same coronal loop, also known as coronal rain limit cycles therefore seem to suggest that the heating is sustained in a quasi-steady manner for hours to days.

In addition to coronal rain often repeatedly forming in the quiescent coronal loops, coronal rain also forms following impulsive one-off events, such as solar flares \citep{jing_2016, scullion_2016} and non-flaring reconnection events \citep{liu_2016, kohutova_2019, mason_2019}. A commonly accepted explanation is that the localised heating responsible for the formation of flare-driven rain is caused by non-thermal electrons accelerated during the flare which hit and heat the chromosphere. However, it seems that the onset of local thermal instability and coronal rain formation needs an additional mechanism besides the electron beam heating \citep{reep_2020}, the main reasons being extremely short duration of the electron-induced heating and the heat deposition site not being sufficiently localised. 

However, most of the observed coronal rain events seem to be of a one-off kind, in the sense that the loop undergoes only 1 cycle of heating and cooling, or is in a complex magnetic field topology involving other loop systems (and reconnection between them). This implies that quasi-constant heating functions are an oversimplification for coronal rain modelling. Importantly, the strong changes in magnetic connectivity expected within an active network of the Sun are completely lacking in the current numerical modelling efforts for coronal rain. Coronal heating is in fact likely to be a strongly variable phenomenon that is impulsive in nature and subject to the continuous and multi-scale perturbations in the photosphere from magneto-convection. If the typical frequency of the individual heating events is much higher than the inverse of the loop cooling timescale, then such heating can be considered quasi-steady. There is however no evidence in general that this assumption about the typical heating frequency is universally valid; for instance, observational evidence suggest that heating in the cores of active regions is highly episodic \citep{testa_2014, reale_2019, testa_2020}. 

Large-scale coronal simulations show that when including the convection zone in the simulation domain, hot chromosphere and corona are self-consistently maintained, with the typical duration of the episodic heating events varying from 2 to 5 minutes \citep{hansteen_2015}. While it is still a subject of debate whether global simulations can realistically model a coronal heating mechanism, it is worth investigating if such self-consistent heating can lead to coronal rain formation. 

All previous numerical studies of coronal rain formation have so far relied on artificial coronal heating terms added to the energy equation, which is usually of the form:

\begin{equation}
 \frac{D \epsilon}{Dt} = - P \nabla \cdot \vec{v}  + \rho \vec{g} \cdot \vec{v} + \nabla \cdot (\vec{\kappa} \cdot \nabla T) + Q_{\mathrm{heat}} - Q_{\mathrm{cool}}  \, ,
\end{equation}

where $\frac{D \epsilon}{Dt}$ is the advective derivative of the energy density, $\vec{\kappa} = \kappa_0 T^{5/2} \hat{\vec{b}}$ is the Spitzer conductivity along magnetic field lines, $Q_{\mathrm{heat}}$ and $Q_{\mathrm{cool}}$ are heating and radiative cooling rates, and $\rho$, $\vec{v}$, $T$, $P$ and $\vec{g}$ are the plasma density, velocity, temperature, pressure and gravitational acceleration respectively. Such user-defined heating term usually has a form of an exponentially decreasing function along the vertical coordinate $y$; $Q_{\mathrm{heat}} = c_0 \exp \Big(-\frac{y}{\lambda}\Big)$, where $c_0$ is the peak heating rate and $\lambda$ is the heating scale height. The user defined heating is therefore highly stratified, spatially smooth and steady \citep[e.g.][]{muller_2003, fang_2013, fang_2015, xia_2017}. The need for the user-defined heating in the previous coronal rain simulations arises from the fact that they typically do not include any self-consistent dissipation mechanisms. They also do not cover complete lower solar atmosphere including chromosphere, photosphere and convection zone, therefore omitting key physical processes in the lower atmosphere, such as magneto-convection, associated magnetic buffeting, braiding and flows. Another drawback of several coronal rain simulations is the commonly used approximation that all of the plasma cooling (i.e. the process essential for modelling the thermal instability and catastrophic cooling) occurs via optically thin radiative losses. This approximation is perfectly valid in the corona but ceases to apply for cool plasma (below temperatures of a few 100 000 K). Such assumption means that regardless whether the radiative loss function is calculated from CHIANTI \citep{dere_2019} or using scaling law approximations \citep[e.g.][] {rosner_1978}, there is a cut-off temperature for the radiative cooling of the plasma condensations. Once electron recombination starts during the cooling process, the energy gained (which depends on the ionisation potential and thus also on the ionisation degree of the plasma) is expected to slow down the cooling rate. The thermal evolution of the plasma condensations is therefore not modelled correctly at low temperatures in the previous coronal rain simulations.

\begin{figure*}
	\includegraphics[width=43pc]{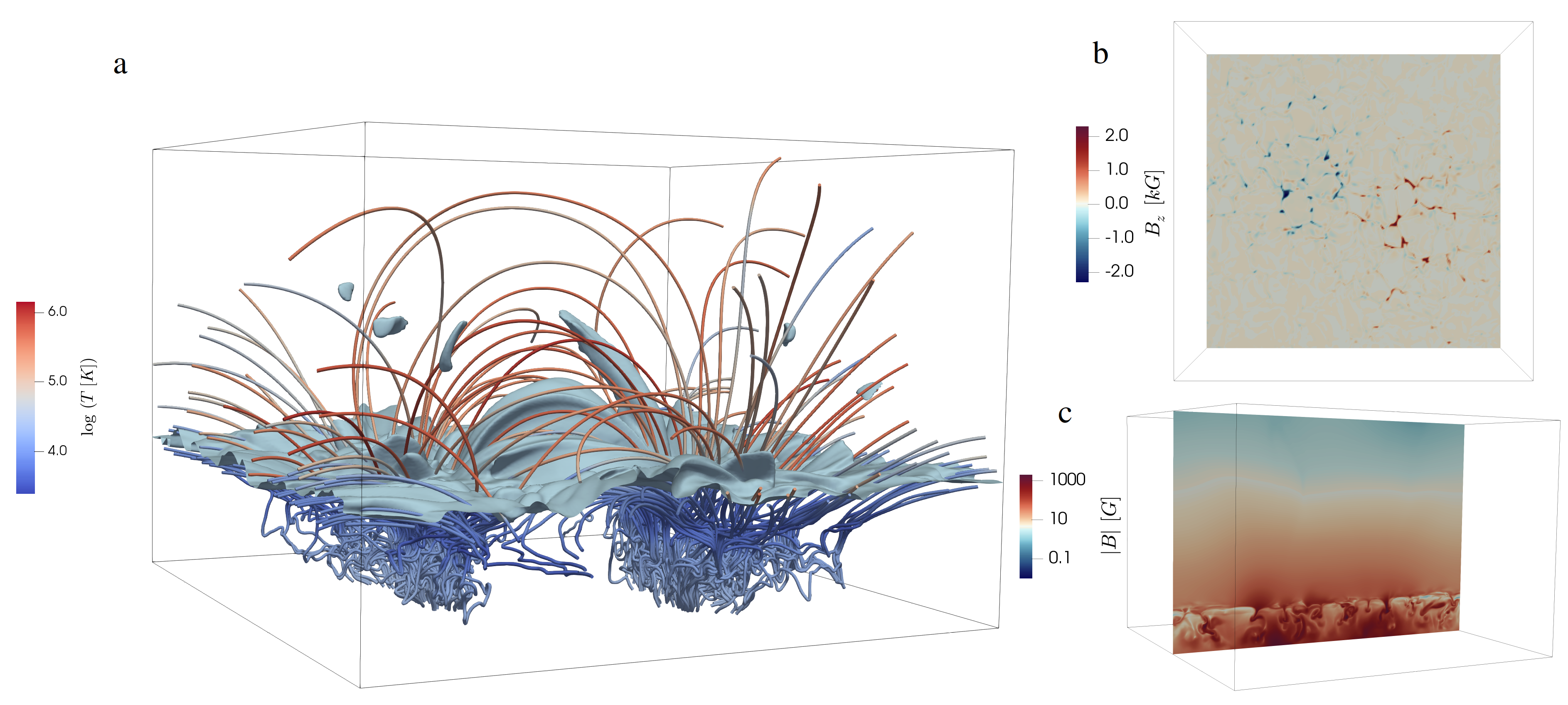}
	\caption{\textbf{a)} Magnetic configuration of the simulation domain with physical size of 24 $\times$ 24 $\times$ 16.8 Mm at $t = 180$ s after the non-equilibrium ionisation of hydrogen has been switched on. The colour of the individual magnetic fieldlines corresponds to their temperature. The $5 \times 10^{-12}$ kg m$^{-3}$ density isosurface is shown in blue. Several cool and dense condensations have formed at coronal heights. \textbf{b)} Line-of-sight component $B_{\mathrm{z}}$ of the photospheric magnetic field at $z = 0$. \textbf{c)} The variation of the of the magnitude of magnetic field strength in vertical direction at $y = 12$ Mm. Animation of this figure is available.}
	\label{fig:context}
\end{figure*}

\begin{figure}
	\includegraphics[width=21pc]{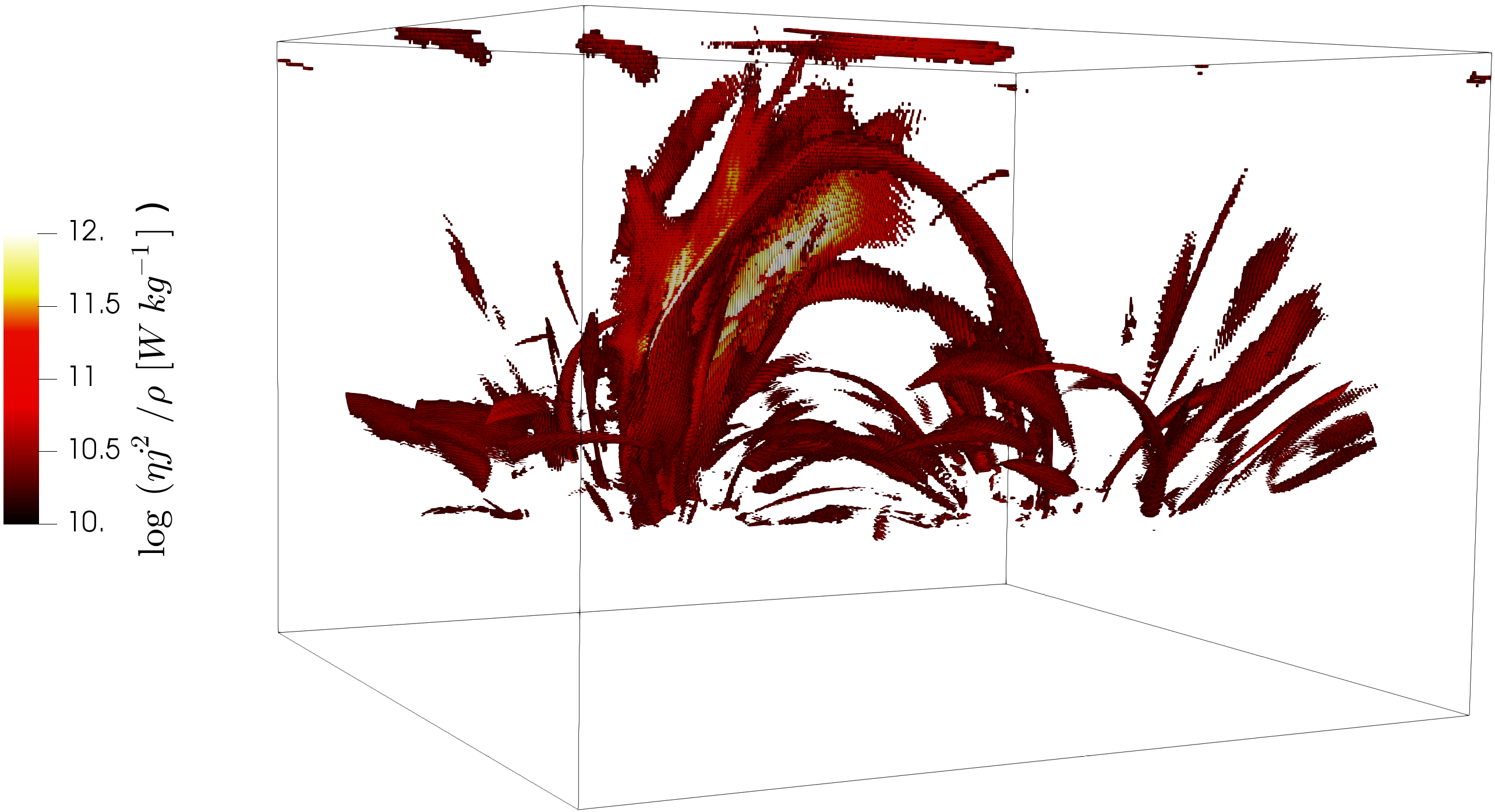}
	\caption{The spatial distribution of Joule heating per unit mass shown for values above the threshold $\eta j^2/\rho=2 \times 10^{10}$ W kg$^{-1}$. Several current sheets are present in the coronal part of the domain.}
	\label{fig:current}
\end{figure}

In order to reproduce large-scale properties of the solar corona including the development of thermal instability in numerical simulations, it is necessary for the nature of the heating in such simulations to be more realistic. One way to do this is to look at large scale response of the corona using complete convection zone to corona simulations. In these models magnetic fields are braided by photospheric and convective motions resulting in the development of current sheets and the associated dissipative heating \citep[e.g.][]{hansteen_2015, kanella_2017}. Furthermore, MHD waves, the other main candidate of coronal heating, are constantly being produced and dissipated throughout the coronal volume. 

In this work we use 3D radiative MHD simulations with Bifrost to self-consistently investigate the development of thermal instability and coronal rain in coronal loops. We also investigate the relation between the spatial distribution of the energy dissipation and the thermal stability of coronal loops and compare this to analytical models. We finally address the relation between the typical duration of the impulsive heating events and cooling timescales of cool plasma condensations.

\begin{figure*}
	\includegraphics[width=43pc]{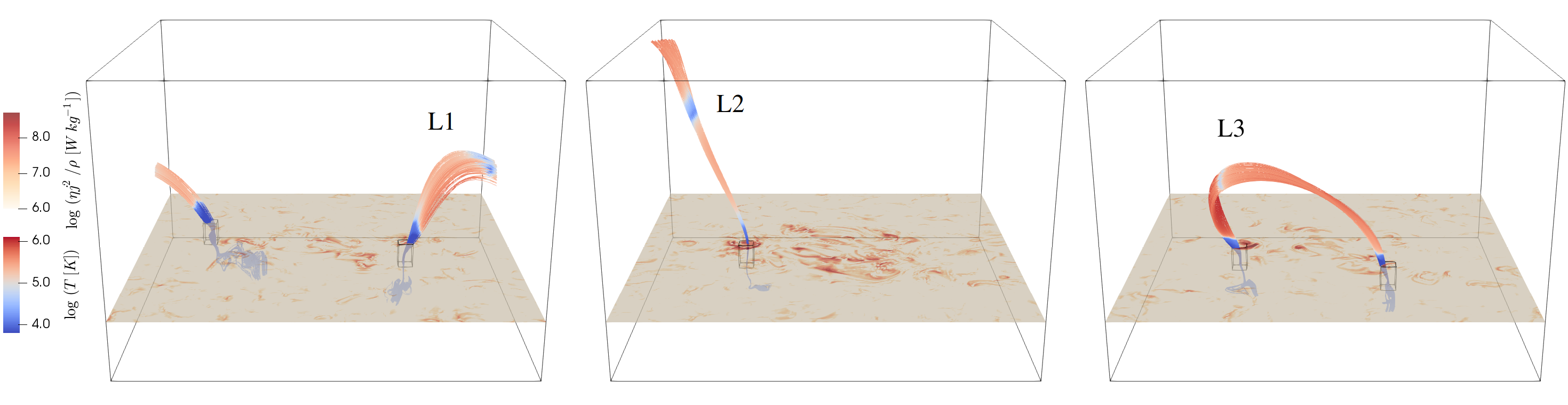}
	\caption{Snapsots of thermally unstable loops L1, L2 and L3 which form cool and dense condensations taken at $t = 1230$ s, $t = 430$ s and $t = 180$ s, respectively. We show 100 field lines which intersect the condensation in each loop, with their colour corresponding to the temperature of the plasma. The surface at $z = 1.2$ Mm shows the concentrations of strong Joule heating in the chromosphere. The regions outlined in black with physical extent of 1 Mm $\times$ 1 Mm $\times$ 1.5 Mm mark the loop footpoint regions in the lower atmosphere.}
	\label{fig:loops}
\end{figure*}

\section{Numerical Model}

For our purpose we use a numerical simulation of a magnetic enhanced network \citep{carlsson_2016} using the 3D radiation MHD code Bifrost \citep{gudiksen_2011}. Bifrost solves resistive MHD equations on a staggered Cartesian grid and includes non-LTE radiative transfer in the photosphere and low chromosphere (the elements included in the radiative transfer calculation are H, He, C, N, O, Ne, Na, Mg, Al, Si, S, K, Ca, Cr, Fe and Ni) and parametrized radiative losses and heating in the upper chromosphere, transition region and corona. The simulation further includes the effects of thermal conduction parallel to the magnetic field and the non-equilibrium ionization of hydrogen in the equation of state.

The simulation is carried out on 504 $\times$ 504 $\times$ 496 grid with physical size of 24 $\times$ 24 $\times$ 16.8 Mm. The grid is uniform in the $x$ and $y$ direction with grid resolution of 48 km. The photosphere corresponds to $z=0$ surface and is defined as the height where the optical depth $\tau_{500}$ is equal to unity (this is an approximation as this height changes slightly over the duration of the simulation). The vertical extent of the grid spans from 2.4 Mm below the photosphere to 14.4 Mm above the photosphere, thus spanning from the upper convection zone to the corona. The vertical resolution is non-uniform in order to resolve steep gradients in density and temperature in the lower solar atmosphere. The grid spacing in $z$ direction varies from 19 km in the photosphere, chromosphere and transition region to 98 km in the upper corona. 

The domain boundaries are periodic in the $x$ and $y$ direction and open in the $z$ direction. The top boundary uses characteristic boundary conditions such that the disturbances are transmitted through the boundary with minimal reflection (\citealt{gudiksen_2011}, appendix A). At the bottom boundary the magnetic field is passively advected while keeping the magnetic flux through the bottom boundary constant (i.e. no additional magnetic field is fed into the domain). Although the staggered grid formulations are inherently magnetic field divergence-free, the numerical round-off errors can accumulate. This cumulative error is handled by the parabolic divergence cleaning every 1000 time-steps.

The average unsigned photospheric magnetic field strength is about 50 G and is concentrated in two patches of opposite polarity about 8 Mm apart in the horizontal plane. This configuration leads to development of several magnetic loops at coronal heights (Fig. \ref{fig:context}).

The simulation is initialised from a hydrodynamic simulation with 3 Mm vertical extend, which is left to relax and then extrapolated in the vertical direction assuming hydrostatic equilibrium to create chromosphere and corona. The large-scale magnetic field configuration was determined using potential field extrapolation from vertical magnetic field specified at the bottom boundary. After the magnetic field has been inserted into the domain it is quickly swept around by convective motions. The non-equilibrium hydrogen ionization is subsequently switched on. 

The high temperature in the chromosphere and the corona is maintained by Ohmic and viscous heating resulting from magnetic field braiding by the convective motions. These are controlled by the numerical resistivity and viscous diffusivity terms respectively. An artificial heating term is switched on for plasma cooling below 2500 K, in order to prevent temperature from reaching too low values in rapidly expanding regions. Aside from relatively few points where this happens, the vast majority of the simulation domain is heated self-consistently via small scale reconnection events that either directly heat the plasma via Ohmic dissipation or induce small scale shear flows that are thermalized via viscous dissipation. The code further employs a diffusive operator which is necessary to maintain numerical stability; this consist of a small global diffusion term as well as of a directionally-dependent hyper diffusion component which enhances the diffusion locally where it is needed the most, e.g. at the shock fronts, while keeping the features sharp elsewhere. Further details of the numerical setup can be found in \citet{carlsson_2016}.

\section{Coronal rain formation and evolution}

\begin{figure*}
\includegraphics[width=43pc]{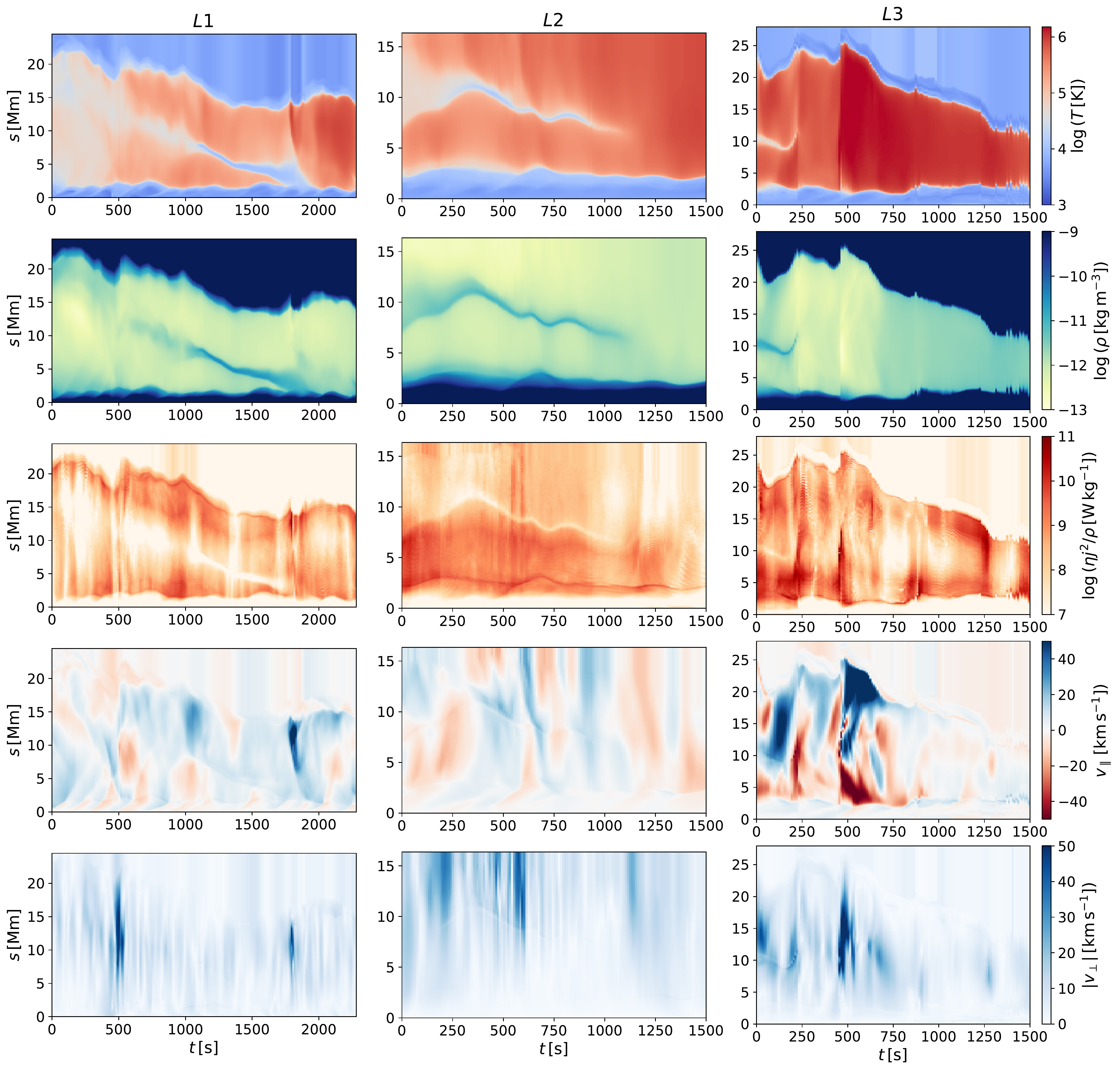}
\caption{Evolution of temperature, density, Joule heating per unit mass, velocity along the magnetic field and the magnitude of the velocity in the plane perpendicular to the magnetic field (top to bottom) along thermally unstable loops L1, L2 and L3 (left to right) following the formation, evolution and potential destruction of the cool plasma condensations. The $x$ axis corresponds to time and the $y$ axis corresponds to the position along the loop measured from left to right footpoint.}
\label{fig:evolution}
\end{figure*}

As the simulated corona is heated by several small-scale heating events, the heating is intrinsically spatially localised and intermittent. The magnetic field is braided by photospheric and convective motions resulting in the development of current sheets at chromospheric and coronal heights \citep[e.g.][]{hansteen_2015, kanella_2017}. The current density in the current sheet then scales as:

\begin{equation}
j = \nabla \times \vec{B} \sim \frac{\Delta B}{\Delta s} \sim \frac{\sin{\Phi} B}{\Delta s} \, ,
\end{equation}

where $B$ is the magnitude of the magnetic field strength in the vicinity of the current sheet, $\Delta s$ is the thickness of the current sheet and $\Phi$ is the angle between the field lines on the opposite of the current sheet \citep{baumann_2013, hansteen_2015}. This then leads to the Joule volumetric heating rate:

\begin{equation}
Q_{\mathrm{Joule}} = \eta j^2 \sim \eta \frac{\sin^2{\Phi} B^2}{\Delta s^2}\, .
\end{equation}

The increased current density in current sheets hence leads to enhanced Ohmic dissipation; they therefore align with regions of enhanced Joule heating (Fig. \ref{fig:current}). Joule heating is strongest in the low chromosphere. However, it should be noted that how much the volumetric heating rate can increase local temperature is dependent on the local plasma density. Joule heating per unit mass (or alternatively \textit{per particle}) is therefore highest in the upper chromosphere, transition region and low corona.

Over the duration of the simulation, several cool and dense condensations can be seen to form at coronal heights. As the thermal conduction is restricted to the direction along the magnetic field, most of the matter and energy transfer occurs along the magnetic field lines. In order to investigate the link between thermal instability occurring in the corona and heating events that can occur anywhere along a fieldline and the associated mass and energy flows, it is necessary to trace the magnetic fields through both time and space. A magnetic fieldline is defined as a curve in 3D space $\vec{r}(s)$ parametrised by the arclength along the curve $s$ for which $\mathrm{d} \vec{r}/ \mathrm{d} s = \vec{B} / |\vec{B}|$. The tracing of the magnetic fieldlines is done by inserting seed points into the locations where the dense condensations can be observed at a given time step, usually shortly after their formation. The seed points are then passively advected both forwards and backwards in time based on the value of the velocity at the seed point position. The spatial coordinates of the traced field line at the given time step are then determined by tracing the magnetic field through the instantaneous seed point position. The accuracy of this method is of course limited by the size of the time step between the two successive snapshots, it however works well provided that the evolution is smooth and there are no large amplitude velocity variations occurring on timescales shorter than the size of the time step.

\subsection{Evolution of thermally unstable fieldlines}

\begin{figure*}
	\includegraphics[width=43pc]{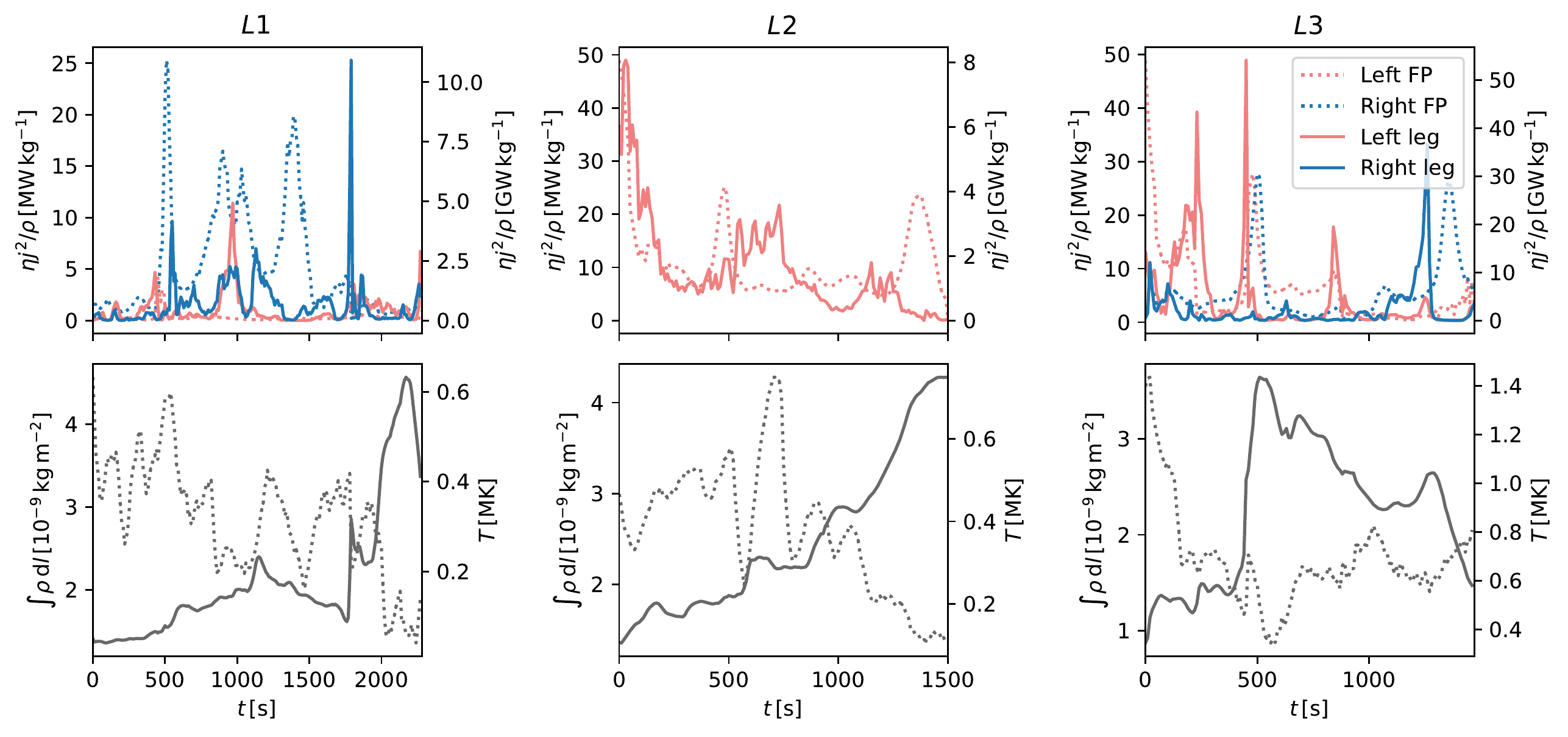}
	\caption{Top: Evolution of the Joule heating at the footpoints of the loops below z = 1.2 Mm in the chromosphere (dotted line, left axis) and 3 Mm above the transition region in the corona (solid line, right axis) for loops L1, L2 and L3. Red and blue plots correspond to the heating evolution at the left and right loop footpoint respectively. Bottom: Evolution of density of the plasma integrated along the coronal portion of the loop (dotted line), and average temperature in the coronal part of the loop (solid line) for L1, L2 and L3.}
	\label{fig:fph}
\end{figure*}

\begin{figure*}
	\includegraphics[width=43pc]{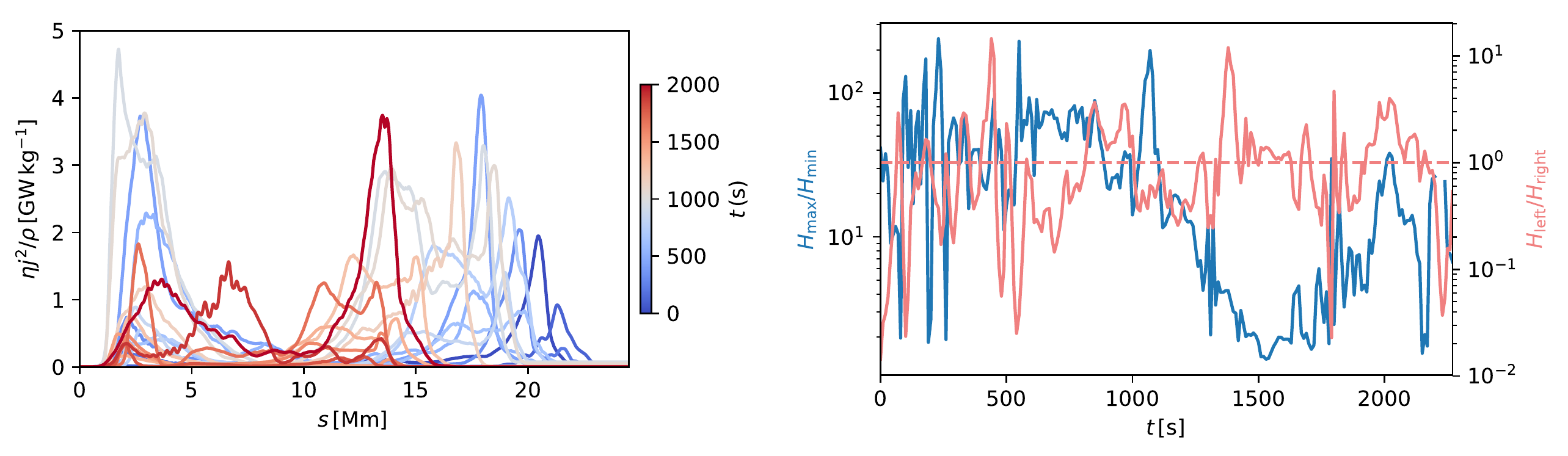}
	\caption{Left: Joule heating per unit mass along the length of the coronal loop L1 plotted every 100 s. The $x$ axis correponds to the position along the loop measured from left to right footpoint. The profiles have been smoothed with a boxcar average with kernel length of 0.5 Mm for clarity. Right: Evolution of the heating asymmetry quantified as the ratio of the maximum heating in the loop legs to minimum heating at the loop apex $H_{\mathrm{max}}/H_{\mathrm{min}}$ (blue) and left-to-right asymmetry in heating of the loop legs $H_{\mathrm{left}}/H_{\mathrm{right}}$ (red) for the loop L1. The red dashed line indicates where $H_{\mathrm{left}}/H_{\mathrm{right}} = 1$, i.e. the line corresponding to perfectly symmetric heating for reference.}
	\label{fig:hl1}
\end{figure*}

We investigate the evolution of the plasma along 3 different fieldlines that show formation of plasma condensations, fieldline marked L1 which intersects the $x$ boundary (the evolution along the fieldline is however continuous as the horizontal boundaries are periodic), fieldline L2 intersecting the upper boundary and fieldline L3 located near the centre of the domain (Fig. \ref{fig:loops}).

The evolution of the temperature, density, Joule heating rate per unit mass and longitudinal and transverse velocity components along each fieldline is shown in Fig. \ref{fig:evolution}. The component of velocity along magnetic field is determined by $v_{\parallel} = \vec{v} \cdot \vec{B}/|\vec{B}|$ (note that positive $v_{\parallel}$ corresponds to the direction along the magnetic field and negative $v_{\parallel}$ correspond to the direction opposing the magnetic field); from this we determine the magnitude of the velocity component in the plane perpendicular to the magnetic field vector as $|v_{\perp}| = \sqrt{ v^2- v_{\parallel}^2}$. In loop L1 a full thermal evolution associated with the formations of condensations is observable. Such evolution includes heating, chromospheric evaporation and increase of plasma density in the loop, onset of thermal instability and formation of plasma condensation which then falls down towards the solar surface and drains the loop. The coronal loop is long-lived and does not undergo any sudden drastic changes. The total length of the coronal loop however slowly decreases from more than 20 Mm to about 10 Mm over the duration of the simulation. Such topological changes can in principle contribute to the loss of the thermal stability of the loop. Longer loops are more likely to become unstable given fixed scale height of the heating \citep{muller_2004, muller_2005}, and a sudden increase in loop length, for example due to magnetic reconnection can result in development of thermal instability \citep{Kaneko_2017}. However, as the reverse scenario is true in the studied loop, it is fair to assume that the decrease in the loop length does not significantly affect the onset of thermal instability. Several impulsive heating events with limited duration can be seen occurring along both loop legs, although they dominate in the left leg ($s = 0$ at the left footpoint). After the condensation falls into the chromosphere at $t = 1780$ s, the loop is evacuated leading to the decrease of the loop density, and subsequently reheated to MK temperatures.

The structure L2 corresponds to an open fieldline. The heating here is more steady than in L1 and does not consist of clearly isolated large magnitude heating events. At the upper boundary the material is allowed to leave the domain with minimum reflection. It is however still possible to accumulate sufficient amount plasma in the upper half of the fieldline necessary to trigger thermal instability and form plasma condensations. This can explain observations of coronal rain along seemingly open magnetic field lines \citep{mason_2019}. The condensation oscillates longitudinally as it falls towards the chromosphere. The evolution of physical quantities along L2 is more gradual and the condensation is slowly reheated back to coronal temperatures shortly before it reaches the chromosphere. We note that after reheating of the condensation, the average loop temperature continues to increase accompanied by a decrease in density of the coronal loop (Fig. \ref{fig:fph}). We attribute this to the fact that this field line is open; the initial expansion caused by the heating at $t \sim 1200$ s leads to an outflow through the upper boundary and hence to the decrease in the overall density of the loop.

The fieldline L3 lies in the centre of the domain. The condensation in L3 has been formed before the non-equilibrium hydrogen ionization has been switched on, the evolution of the fieldline prior to condensation formation is therefore not studied. The condensation was destroyed by a reconnection event occurring in the same loop at $t=240$ s, when a clear jump in the fieldline evolution can be seen due to the seed point advection breaking down. This is accompanied by a sudden onset of strong Joule dissipation in the large fraction of the left loop leg, lasting for 100 s. This lead to the condensation being rapidly reheated back to coronal temperatures. Following the destruction of the condensation, L3 is subject to several additional reconnection events identifiable as short bursts of strong Joule heating. A dramatic increase in the average loop temperature occurs following one such event at $t \sim 500$ s (Fig. \ref{fig:fph}). The resulting bidirectional outflows along the loop lead to a brief decrease in the loop density; the evolution however quickly reverses, as the resulting footpoint heating triggers evaporative flows which steadily raise the loop density. The heating rates per unit mass associated with such reconnection events reach $10^{11}$ W kg$^{-3}$, which is at least an order of magnitude higher than in L1 and L2. As a result, L3 is heated to higher temperatures than the two other loops studied in this work.

\subsection{Response of loops to heating events}

\begin{figure}
	\includegraphics[width=20pc]{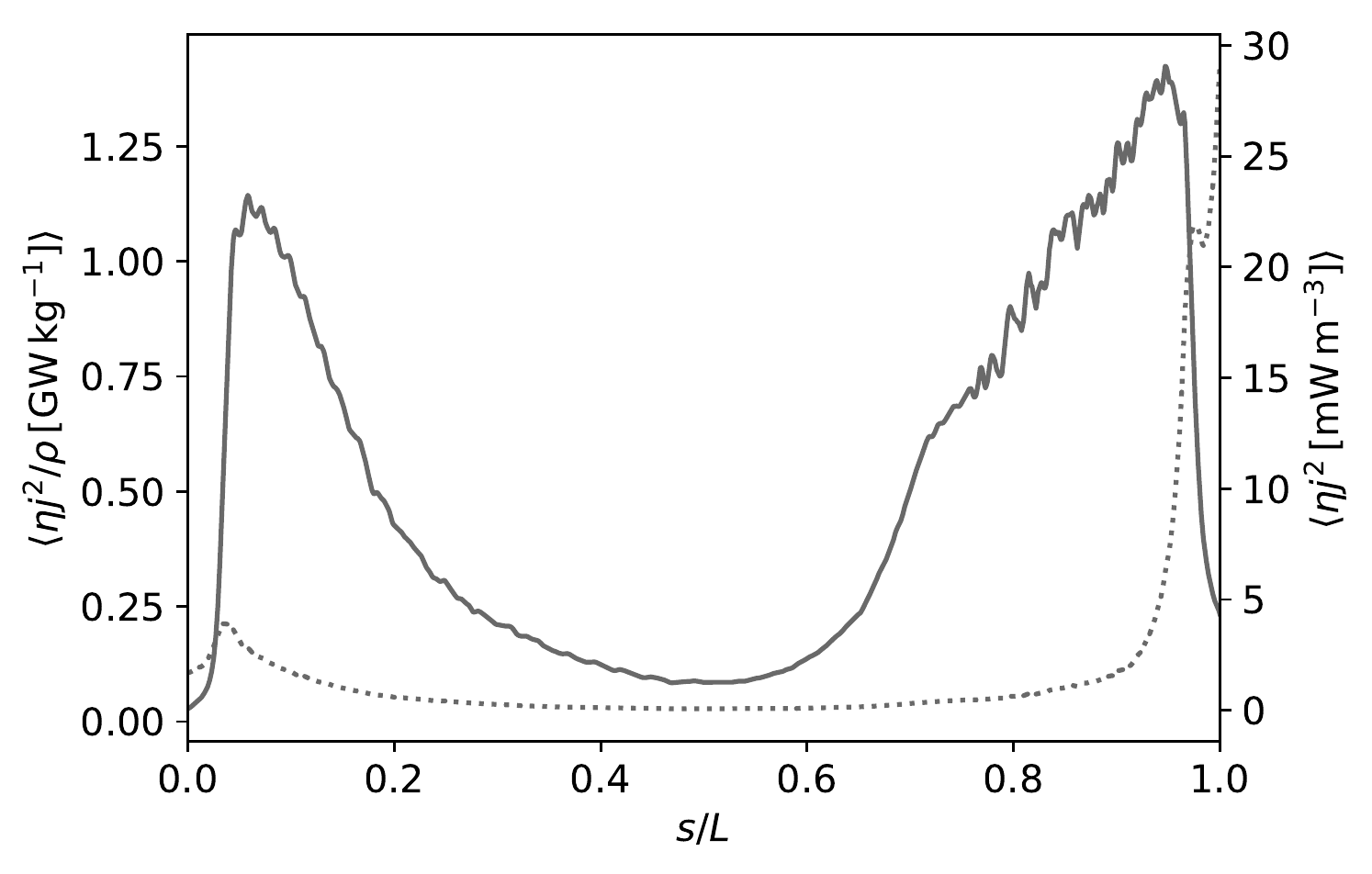}
	\caption{Time-averaged heating per unit mass (solid line) and volumetric heating rate (dotted line) along loop L1 normalised by loop length.}
	\label{fig:avg_heating}
\end{figure} 

The footpoints of the thermally unstable fieldlines, with the exception of the left footpoint of L1, are embedded in the regions of enhanced Joule heating at $z = 1.2$ Mm (Fig. \ref{fig:loops}).

The evolution of the Joule heating along L1 shows two major impulsive heating events in the left loop leg, at $t = 320$ s and $t = 820$ s lasting 150 and 200 s respectively (Fig.~\ref{fig:evolution}). The enhanced heating associated with these events has large spatial extent along the loop spanning considerable fraction of the loop length. Other than these two impulsive events with limited duration, there is almost no sustained  enhanced heating concentrated at the very bottom of the loop in the left loop leg. In right loop leg however, the enhanced heating is sustained for extended periods of time and is mostly concentrated in the transition region close to the footpoint before and during the condensation formation. The scale height of the right loop leg heating is very short, leading to more concentrated heating than for the left loop leg. During the second impulsive heating event at $t = 820$ s, an enhancement in the right loop leg heating can be observed as well. There is another reconnection event occurring at $t = 1800$ s in the right loop leg as suggested by a discontinuity in the field line evolution and associated short burst in the Joule heating. Following each impulsive heating event an upflow of the material into the loop from the lowermost part of the loop can be observed (Fig. \ref{fig:evolution}). These collectively contribute to the increase of the density of the plasma in the upper parts of the loop.  

Unlike L1, the fieldline L2 is subject to heating with gradually decreasing strength with the enhanced heating phase lasting nearly 1000 s (Fig. \ref{fig:evolution}). The heating is concentrated in the transition region and lower corona. No large amplitude impulsive heating events at transition region/coronal heights take place during the lifetime of the structure. We instead conclude that the magnetic stress leading to the dissipative heating is gradually build up in the fieldline as it is gradually twisted and braided together with the surrounding closed loops. Their presence is crucial, as it would be very difficult to build up sufficient amounts of magnetic stress in an isolated open fieldline bundle. In such case the magnetic twist would rapidly propagate away with the local Alfv\'{e}n speed.

We also investigate how the evolution of the heating at the loop footpoints in the chromosphere correlates with the heating in the transition region and lower corona and the density in the coronal part of the loop. Figure \ref{fig:fph} shows the evolution of the mean Joule heating rate per unit mass integrated over a region of the spatial extent of 1 Mm $\times$ 1 Mm $\times$ 1.5 Mm covering the chromospheric footpoints of each loop below $z = 1.2$ Mm (the footpoint regions of all loops are shown in Fig. \ref{fig:loops}). It also shows the corresponding evolution of the temperature and density in the coronal part of the loop (i.e. above the transition region and including the condensations). Impulsive heating events in the footpoints can lead to a delayed increase in the amount of plasma integrated along the coronal fraction of the loop. However, in several cases it is not possible to establish one-to-one correspondence between the footpoint heating events and the loop density increases, as the timescales for the transport of plasma from the chromosphere into the corona vary with local physical quantities and conditions. A lower bound of the time scale for the transport of evaporated plasma into the corona can be determined using the estimate for the average sound speed in each loop (assuming the flows are subsonic). The sound speed is given by $c_{\mathrm{s}} = \sqrt{\gamma \bar{p}/ \rho}$ where $\bar{p}$ is the average pressure in the loop. We obtain 25 km s$^{-1}$, 46 km s$^{-1}$ and 83 km s$^{-1}$ leading to time scales of 200 s,  140 s and 60 s for loop L1, L2 and L3 respectively, assuming the plasma is transported 5 Mm up into the corona. Also, as pointed out earlier, on multiple occasions the enhanced heating is localised higher in the loop legs at coronal height and has no counterpart in the chromospheric footpoints. This is the case for the left footpoint of loop L1, where despite several instances of enhanced impulsive heating observed in the loop legs there is negligible footpoint heating observable below $z = 1.2$ Mm. Enhanced heating at the footpoints is therefore not a prerequisite for a large fraction of the coronal loop to be heated.  

We further focus on the evolution of the heating profile along the thermally unstable fieldline L1 as it captures a full thermal evolution. The evolution of the Joule heating profile along the length of the coronal loop is shown in Fig. \ref{fig:hl1}. The spatial extent of the regions along the loop showing enhanced Joule heating varies from 2 Mm up to 8 Mm, which is a significant fraction of the total length of the loop. We estimate the heating scale height from the time-averaged heating profile along the loop to be 20\% of the total loop length (Fig. \ref{fig:avg_heating}); this value is however only a rough estimate given the large variability of the individual heating events. During the entire loop evolution heating in the loop legs always dominates over the heating at the apex of the loop.  

\begin{figure*}
	\includegraphics[width=43pc]{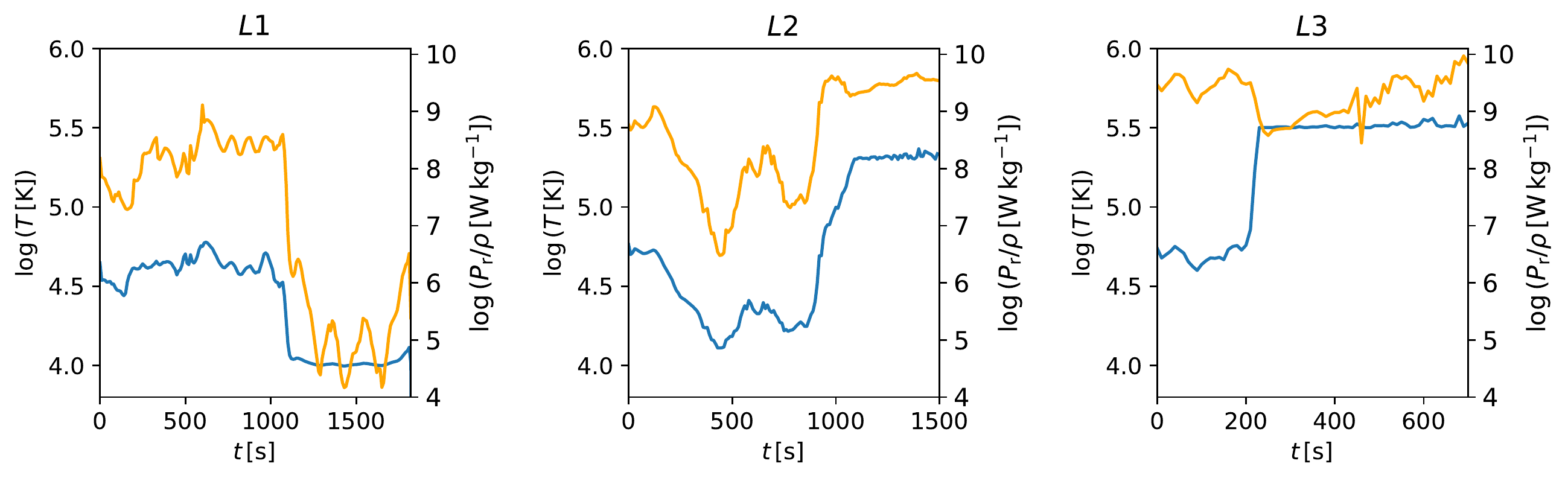}
	\caption{Left to right: Evolution of loop minimum temperature in the corona (blue) and corresponding optically thin radiative loss rate per unit mass (orange) for loops L1, L2 and L3.}
	\label{fig:cooling}
\end{figure*} 

We quantify the asymmetry of the loop heating with respect to height as $H_{\mathrm{max}}/H_{\mathrm{min}}$. Here $H_{\mathrm{max}}$ is the average heating rate per unit mass in the loop leg segment with 2 Mm length centred on the location with maximum Joule heating along the loop and $H_{\mathrm{min}}$ is the average heating rate per unit mass in the loop apex segment with 5 Mm length centred on the location with minimum Joule heating. During the first 1000 s of the loop evolution, $H_{\mathrm{max}}/H_{\mathrm{min}} > 10 $ during sustained periods and the heating is localised in the lower part of the loop (Fig. \ref{fig:hl1}). After the condensation is formed in the loop, this is no longer the case; the heating distribution along the loop is much more uniform and the heating close to the loop apex is comparable to the heating in the loop legs. This requirement on the ratio between the maximum and minimum heating for the loop to become thermally unstable is in good agreement with estimates from an analytical model of a loop subject to stratified heating \citep{klimchuk_2019}:
\begin{equation}
\frac{H_{\mathrm{max}}}{H_{\mathrm{min}}} > \Bigg( 1 + \frac{c}{\Gamma}\Bigg)
\end{equation}
Where $c = r_{\mathrm{tr}}/r_{\mathrm{c}}$ is the ratio of the radiative losses in the transition region and corona and $\Gamma = A_{\mathrm{c}}/A_{\mathrm{tr}}$ is the loop expansion factor. Assuming $c \sim 10$ and $\Gamma \sim 1$ which is valid for a short loop with small area expansion, the critical value for the instability onset $H_{\mathrm{max}}/H_{\mathrm{min}} \sim 11$.

Similarly, we quantify the asymmetry of the heating between the left and right part of the loop by $H_{\mathrm{left}}/H_{\mathrm{right}}$, i.e. as a ratio of the average heating in the region of 2 Mm longitudinal extent centred on the position of the maximum heating in the left and right loop leg respectively. During the initial stages of the loop evolution, $H_{\mathrm{left}}/H_{\mathrm{right}} < 2$ during most of the loop evolution except for the impulsive heating events at $t = 320$ s and $t = 820$ s where $H_{\mathrm{left}}/H_{\mathrm{right}} \sim 10$ and $ \sim 2$ respectively (Fig. \ref{fig:hl1}). The left-to-right heating asymmetry is therefore strongest during the periods of enhanced heating which drive the onset of the catastrophic cooling and the condensation formation. The instances where the left-to-right heating asymmetry is large, such that $H_{\mathrm{left}}/H_{\mathrm{right}} > 5$ or $H_{\mathrm{left}}/H_{\mathrm{right}} < 0.2$ have very short duration of less than 50 s, as resulting large pressure differences along the loop are rapidly equalised by internal flows. The overall left-to-right asymmetry is also consistent with the analytical estimate of $H_{\mathrm{left}}/H_{\mathrm{right}} > 3$ as the threshold on the heating asymmetry for development of thermal instability \citep{klimchuk_2019}. A persistent asymmetry that is greater than a threshold value will instead lead to an onset of unidirectional siphon flow between the loop footpoints \citep{patsourakos_2004, xia_2011, klimchuk_2019}. 

\subsection{Cooling of condensations}

We investigate the evolution of temperature and optically thin radiative loss rate per unit mass (Fig. \ref{fig:cooling}). Following the catastrophic cooling phase, the condensations in L1 and L2 cool down to $\sim$ 10 000 K. The condensation in L3 on the other hand has a steady temperature of ~ 50 000 K before being reheated to coronal temperatures. The timescales for catastrophic cooling to chromospheric temperatures vary from 150 s for condensation formed in loop L1 where the condensation formation is triggered by impulsive heating event to 400 s for the condensation in L2, where the heating is more gradual and sustained for extended period of time. The evolution suggest that the cooling profiles are not a simple exponential or two-step profiles inferred from the observations \citep{antolin_2015}. This is especially true in the case of the condensation formed in L2 that is subject to more gradual cooling and which is briefly reheated following the initial cooling stage.

We estimate the radiative cooling timescales of the 3 loops based on the average values of density $\bar{\rho}$ and temperature $\bar{T}$ in the coronal part of the loops. The radiative cooling timescale is given by 
\begin{equation}
\tau_{\mathrm{rad}} \sim \Bigg(\frac{2}{\gamma - 1} \Bigg) \frac{\bar{m} k_{\mathrm{B}} T^{1-\alpha} }{\chi \bar{\rho}} \,, 
\end{equation}
where $\gamma = 5/3$, $k_B$ is the Boltzmann constant, $\bar{m}$ is the mean particle weight and $\alpha$ and $\chi$ are coefficients that approximate the radiative loss function $\Lambda(T) = \chi T^{\alpha}$ as a piece-wise power law \citep{rosner_1978}. We obtain radiative cooling timescales of $\sim$ 130 s, $\sim$ 100 s and $\sim$ 4600 s for loops L1, L2 and L3 respectively; the long cooling timescale in the L3 case is due to the loop being hotter and less dense than the other two. These values are in agreement with the evolution of the temperature shown in Fig. \ref{fig:cooling}, aside for the condensation in L3, formation of which we do not analyse as it occurs before non-equilibrium hydrogen ionisation is switched on in the simulation. It should be noted that these timescales should be treated as an order-of-magnitude estimates only, as they are strongly dependent on the plasma density which changes during the catastrophic cooling.

\section{Discussion and conclusions}

We have for the first time studied the formation and evolution of coronal rain condensations in a 3D simulation of the solar atmosphere spanning from convection zone to corona, which correctly models the chromosphere by including non-LTE radiative transfer and non-equilibrium ionisation of hydrogen and in which the atmosphere is self-consistently heated through magnetic field braiding. This ensures a realistic response of the chromosphere to the impulsive heating events and that cooling of the condensations is modelled correctly for the full range of temperatures ranging from coronal to chromospheric. The formation of the coronal rain is also for the first time seen in a realistic 3D magnetic field configuration, as the photosperic magnetic field consists of two opposite polarity patches while including the small-scale variability outlining the edges of intergranular lanes, with the overall structure very similar to magnetograms inferred from the solar observations. In addition, the magnetic field strength at coronal heights is of the order of 10 G, i.e. an order of magnitude greater than previous 3D coronal rain simulations \citep{moschou_2015, xia_2017}, more in agreement with typical values in the active region coronal loops \citep{aschwanden_2005}. The solar atmosphere in the simulation is heated self-consistently via dissipation associated with magnetic reconnection and magnetic field braiding. This means that, besides the numerical treatment of diffusivities in Bifrost (appropriate to a global 3D MHD simulation), there are no artificially imposed constraints on the spatial and temporal characteristics of heating. The self-consistent heating is found to be on average spatially localised in the lower atmosphere and impulsive.

The above analysis addresses a number of open questions. Firstly, it is necessary to ask whether impulsive heating events, such as Ohmic dissipation associated with magnetic field braiding can trigger coronal rain formation in the first place. The evolution of loop L1 shows that this is indeed possible, as the duration of the impulsive heating events occuring in the loop is 100 - 200 s, similar to the loop radiative cooling timescale of $\sim$ 100 s. An impulsive heating event of significant amplitude and with duration that is comparable to the radiative cooling timescale is sufficient to cause sufficient chromospheric evaporation into the loop for the loop to become thermally unstable \citep{johnston_2019}. Previous studies that addressed prominence formation with using a time-variable coronal heating term found that formation of condensations still occurs if 1. the heating is intermittent, but the frequency of the individual heating events exceeds the inverse of the radiative cooling timescale \citep{karpen_2008} and 2. if the heating ceases during the condensation formation, the condensation still continues to grow \citep{xia_2011}. It should also be noted that such impulsive events can also act as a perturbation that trigger the catastrophic cooling in a marginally stable plasma. We noted differences between condensations formed in loops L1 and L2, in particular that the heating driving the condensation formation in L1 has an impulsive character compared to the gradual heating of L2. Also, the cooling timescale of the condensation formed in L1 is much shorter than that of the condensation formed along L2, suggesting different conditions at coronal heights. Secondly, we also addressed the issue of the apparent coronal rain formation along open field lines and shown that flows of evaporated plasma can lead to sufficient density increase to trigger radiative instability in both closed and open magnetic field configurations. Finally, our work also addresses a possible mechanism responsible for destruction of coronal rain condensations before they reach the solar surface. A reconnection event occurring in the upper part of the leg of loop 3 leads to the condensation being heated to very high temperatures and subsequently destroyed as a result. This suggests that the lifetime of the condensations can be dependent on the frequency of such reconnection events in shorter loops. Also, it might possibly explain why coronal rain is not observed in short low-lying loops close to the active region core that are heated to very high temperatures. Even if the condensations do form in such loops (which is significantly less likely as it is easier to trigger thermal instability in longer loops, see e.g. \citet{muller_2004, muller_2005}), they are most likely short-lived.

There are therefore several implications for the conclusions one can draw about spatial and temporal localisation of coronal heating from the coronal rain observations. Compared to user-defined heating used in previous simulations of coronal rain formation, the regions of enhanced heating in the self-consistently heated simulation are not as localised but instead have finite spatial extent along the loop reaching up to 25 \% of the total loop length. Despite this lack of localisation at the very footpoints of the loops, the dissipative heating in the loop legs is nevertheless stronger than at the loop apex. Similarly, because of the nature of the impulsive heating events over the duration of the simulation, left-to-right heating asymmetry between the footpoints is always present to a certain extent, but does not typically exceed the threshold established by \citet{klimchuk_2019}, except for very short periods of time. This is partially a consequence of magnetic field strengths being similar in the two loop footpoints. We note that if this was not the case, the resulting strong asymmetry between the footpoints would instead lead to a unidirectional flow, which may prevent formation of cool condensations.

The highest heating per particle is localised at intermediate heights in the coronal loop legs, rather than at the coronal loop footpoints. This is clearly visible e.g. in the left part of the loop 1, where the Joule heating of the chromospheric part of the left footpoint is negligible; however, at the same time the loop is subject to significant heating higher up in the left loop leg. The footpoint localisation of the heating is often assumed as a condition for development of thermal instability and coronal rain formation. Here we have however shown that this assumption should be relaxed as 1. the Joule heating itself is more efficient in the upper chromosphere, transition region and low corona and 2. heating extended along loop legs is capable of triggering sufficient evaporation into the loop leading to formation of plasma condensations at coronal heights. Heating generated during such an impulsive event is likely redistributed by thermal conduction, leading to both spatially extended regions of enhanced heating along the loop and to the local heating of the chromosphere which produces evaporative flux into the loop.
 
An important point that should not be overlooked is that even though the heating of the upper chromosphere and the corona in the simulation is self-consistent as such, it relies on the prescription of the Ohmic dissipation using magnetic resistivity, which is technically a parametrisation in itself. The actual dissipation processes operate on kinetic scales which cannot be represented by MHD fluid models, and therefore require using a kinetic formulation, modelled in a majority of simulations by a particle-in-cell (PIC) method. To reproduce physical mechanisms responsible for the dissipation and the associated response of the solar atmosphere from the first principles, a coupled MHD-PIC treatment is necessary.

One of the limitations of the presented work is the limited spatial extent of the simulation domain. The simulation only captures the evolution of the corona up to 14.4 Mm, our study is therefore limited to relatively short low-lying loops. This makes comparison with observations difficult, as most of the observational works address coronal rain formation in loops with lengths of at least 100 Mm or longer. This is likely to affect the comparison of quantities which are expected to scale with the length of the loop, such as cooling and heating timescales. The natural next step is therefore extending such simulations into larger spatial scales to study the evolution of coronal loops with lengths of the order of 100 Mm.

Given the required resolution especially in the lower solar atmosphere and the complexity of the individual modules necessary for including non-ideal MHD effects and radiative transfer, there are considerable computational limitations to this. These can be overcome by utilizing adaptive mesh refinement and task-based code frameworks \citep{nordlund_2018} in the Bifrost code, which is an ongoing effort.

Another potentially strong limitation is the limited spatial resolution. \citet{bradshaw_2013} and \citet{johnston_2019} show that in order to properly model the transition region response to coronal heating a vertical spatial resolution down to 2 km is necessary. Underresolving the transition region limits the evaporative flux into the corona which in turn limits the amount of coronal rain that forms in the simulation. 

Self-consistent, global 3D MHD simulations such as this one show that magnetic field braiding leading to magnetic reconnection in the lower atmosphere is common \citep{gudiksen_2005, kanella_2017}. A strong implication of this is the large variability in the magnetic connectivity within an active region, and even within a loop bundle, as demonstrated in this work. We have shown that this connectivity can strongly impact on the formation and evolution of condensations, which naturally places doubt on the existence of long lasting TNE cycles. However, the existence of such cycles, lasting even a week or more, has been observationally demonstrated. This therefore constitutes a huge numerical challenge and conundrum for global 3D MHD simulations \citep{antolin_2020}. A proper resolution of this issue can only come through the extension of these global models to properly take into account evolution of long loops over long time scales. This will determine how realistic our current self-consistent simulations of solar atmosphere are and how well we can reproduce the long term variability of the corona.

\begin{acknowledgements}
This research was supported by the Research Council of Norway through its Centres of Excellence scheme, project no. 262622. P.A. acknowledges funding from his STFC Ernest Rutherford Fellowship (No. ST/R004285/1). This work has received support from the International Space Science Institute, Bern, Switzerland to the International Teams on ‘Implications for coronal heating and magnetic fields from coronal rain observations and modeling’ (PI: P. Antolin) and ‘Observed Multi-Scale Variability of Coronal Loops as a Probe of Coronal Heating’ (PIs: C. Froment and P. Antolin). 
\end{acknowledgements}

\bibliography{rain}

\end{document}